\documentclass[twocolumn]{aastex631}

\usepackage[normalem]{ulem}
\shorttitle{Solar cycle variability vs dynamo supercriticality}
\shortauthors{Ghosh et al.}

\newcommand{\Fig}[1]{Figure~\ref{#1}}
\newcommand{\Figs}[2]{Figures~\ref{#1} and \ref{#2}}

\newcommand{\Eq}[1]{Equation~(\ref{#1})}

\newcommand{\Tab}[1]{Table~\ref{#1}}

\def\bl{Babcock--Leighton}

\begin{document}

\title{Characterizing the solar cycle variability using nonlinear time series analysis at different amounts of dynamo supercriticality: Solar dynamo is not highly supercritical}

\correspondingauthor{Bidya Binay Karak}
\email{karak.phy@iitbhu.ac.in}

\author[0000-0002-8764-5228]{Aparup Ghosh}
\affiliation{Department of Physics,  
University of Massachusetts Amherst,  
Amherst, MA 01003, USA}
\affiliation{Department of Physical Sciences, 
Indian Institute of Science, Education and Research Kolkata, 
Mohanpur 741246, India}

\author{Pawan Kumar}
\affiliation{Department of Physics,  
Indian Institute of Technology (BHU), 
Varanasi 221005, India}

\author{Amrita Prasad}
\affiliation{Department of Data Science, 
Cardiff School Of Technologies, Cardiff Metropolitan University, 
Cardiff - CF5 2YB, UK}

\author[0000-0002-8883-3562]{Bidya Binay Karak}
\affiliation{Department of Physics,  
Indian Institute of Technology (BHU), 
Varanasi 221005, India}



\begin{abstract}
The solar dynamo is essentially a cyclic process in which the toroidal component of the magnetic field is converted into the poloidal one and vice versa.  This cyclic loop is disturbed by some nonlinear and stochastic processes mainly operating in the toroidal to poloidal part.  Hence, the memory of the polar field decreases in every cycle.  On the other hand, the dynamo efficiency and, thus, the supercriticality of the dynamo decreases with the Sun's age.  Previous studies have shown that the memory of the polar magnetic field decreases with the increase of supercriticality of the dynamo.  In this study, we employ popular techniques 
of time series analysis, namely, compute Higuchi's fractal dimension, Hurst exponent, and Multi-Fractal Detrended Fluctuation Analysis, to the amplitude of the solar magnetic cycle obtained from dynamo models operating at near-critical and supercritical regimes.  We show that the magnetic field in the near-critical regime is governed by strong memory, less stochasticity, intermittency, and breakdown of self-similarity.  On the contrary, the magnetic field in the supercritical region has less memory, strong stochasticity, and shows a good amount of self-similarity.  Finally, applying the same time series analysis techniques in the reconstructed sunspot data of 85 cycles and comparing their results with that from models, we conclude that the solar dynamo is possibly operating near the critical regime and not too much supercritical regime.  Thus Sun may not be too far from the critical dynamo transition.
\end{abstract}



\section{Introduction} \label{sec:intro}

The solar cycle is not regular; the amplitude of the cycle has a considerable amount of variation. 
This is best seen in the observed sunspot number plot for the last 400~years and its proxy, such as the concentration of the cosmogenic isotopes ($^{14}$C and $^{10}$Be) data for the last several thousands of years \citep{Uso17, Biswas23}.
Another prominent feature of the cycle irregularity is the Gnevyshev-Ohl/Even-Odd rule which is an alternating pattern of strong and weak cycles \citep{GO48, Hat15}. 
While in some cycles, the amplitude changed drastically from one cycle to the next (e.g., Cycles 6, 20),  there is a reasonably smooth variation as seen by some envelopes in the amplitude, e.g., Gleissberg cycle \citep{glessberg}. This indicates that although the solar cycle is irregular, there is some (temporal) memory in the underlying system.

We have good support that an $\alpha\Omega$ type dynamo model, operating in the solar convection zone (SCZ), is responsible for causing the solar magnetic cycle \citep{Kar14a, CS15, Char20}.
In this type of dynamo the poloidal field (which is observed to become strongest near the solar minimum) gives rise to the toroidal field and thus the sunspots for the next cycle.  Hence, there is an unavoidable memory of about 5 years in the solar dynamo.
This is indeed observed in the observations because there is a strong correlation between the polar field (or its proxy) at the solar minimum (or the peak field/proxy) and the amplitude of the next sunspot cycle \citep{JCC07, WS09,  KO11, Priy14, Pawan21}. In fact, this is true in any $\alpha\Omega$ type 
dynamo model as long as the poloidal field gives rise to the toroidal field \citep{CB11}. The polar precursor method of the solar cycle prediction is indeed based on this idea \citep{Sch78, CCJ07, Pawan21, Bhowmik+Nandy, Pawan22, biswas23mnras}.

Now, is the memory of the polar field limited to the 
next cycle only, or is it propagated to multiple following cycles? At first glance, one may think that the memory should be propagated to multiple cycles as the solar dynamo is just the oscillation between the two components of the magnetic field: poloidal and toroidal.
However, if we carefully analyze the dynamo chain, 
then we find that the toroidal to poloidal part of the solar dynamo is the one that involves some randomness 
arising due to the distributions of the BMR properties, 
primarily due to 
scatters around Joy's law and randomness in the BMR  emergences \citep{JCS14, KM17, KM18, Nagy2017-io, Kar20}.
Hence, in every cycle during the generation of the poloidal field, 
the memory of the polar field is degraded. 
The toroidal to poloidal component of the solar dynamo also involves some nonlinearities, which at least include the flux loss due to magnetic buoyancy in the formation of BMR \citep{Biswas22prl}, latitude quenching \citep{J20, Kar20}, and tilt quenching \citep{Jha20}. 
The nonlinearity plays an important role in determining the memory of the polar field when the dynamo becomes supercritical. \citet{Pawan21b} have shown that if the dynamo operates near the critical dynamo transition, then the dynamo tends to be linear and the polar field of a cycle strongly correlates to the toroidal field of multiple following cycles. On the other hand, this correlation is limited to one cycle if the dynamo operates in a highly supercritical region. They have made this conclusion by performing various types of dynamo simulations at different parameters, namely diffusivity, meridional circulation, and nonlinearity and in all models, they found that the correlation between the polar field and the toroidal field of the subsequent cycles is consistently shortened from multiple cycles to one cycle as the supercriticality of the dynamo is increased.
They have further shown that this degradation of memory is independent of the amount of diffusion and the advection of the magnetic field, as suggested by \citet{YNM08}. The diffusion and advection, of course, determine the memory within a cycle and a little bit beyond one cycle \citep{CD00, JCC07}. However, when these are kept constant, whether the memory of the polar field is propagated to multiple cycles or not is determined by the amount of the supercriticality.  

While in the previous work \citep{Pawan21b}, the memory was measured just by measuring the correlation between the peaks of the polar fields and the peaks of the following cycle toroidal fields. In the present work, we shall apply some 
 well-known 
techniques of nonlinear time series analyses.
In the time series analyses, two popular quantities, namely the Higuchi's dimension ($D$) and Hurst exponent ($H$) are generally used to determine the complexity of the system. $D$ is a fractal dimension that determines the geometrical structure at multiple scales. 
On the other hand, $H$ helps to identify the presence of long-term memory in a time series \citep{RO98, MA07, SZ17, Das22}.

In the present study, we shall compute $D$ and $H$ 
of the cycles produced in the different regimes of the dynamo to measure the memory of the cycles.  Using these measures, we shall independently demonstrate that the 
memory of the solar cycle beyond one cycle is indeed determined by the supercriticality of the solar dynamo. 
We shall also compare the results with those from the observed sunspot data and comment on the supercriticality of the solar dynamo. 
Knowing the amount of supercriticality will tell how far our Sun is from the critical dynamo transition (below which the dynamo action ceased). Answering this is essential because, with age, the Sun's dynamo efficiency decreases (as the rotation rate and thus 
 the generation of poloidal field reduces), 
and at some point in its life, the Sun will stop producing its (large-scale) dynamo action. Incidentally, some studies hint that our Sun is not too far from the critical dynamo transition \citep{R84, Met16, KN17, V23}.

\section{Model data} \label{sec:models}

\begin{figure*}
    \centering
    \includegraphics[width=\textwidth]{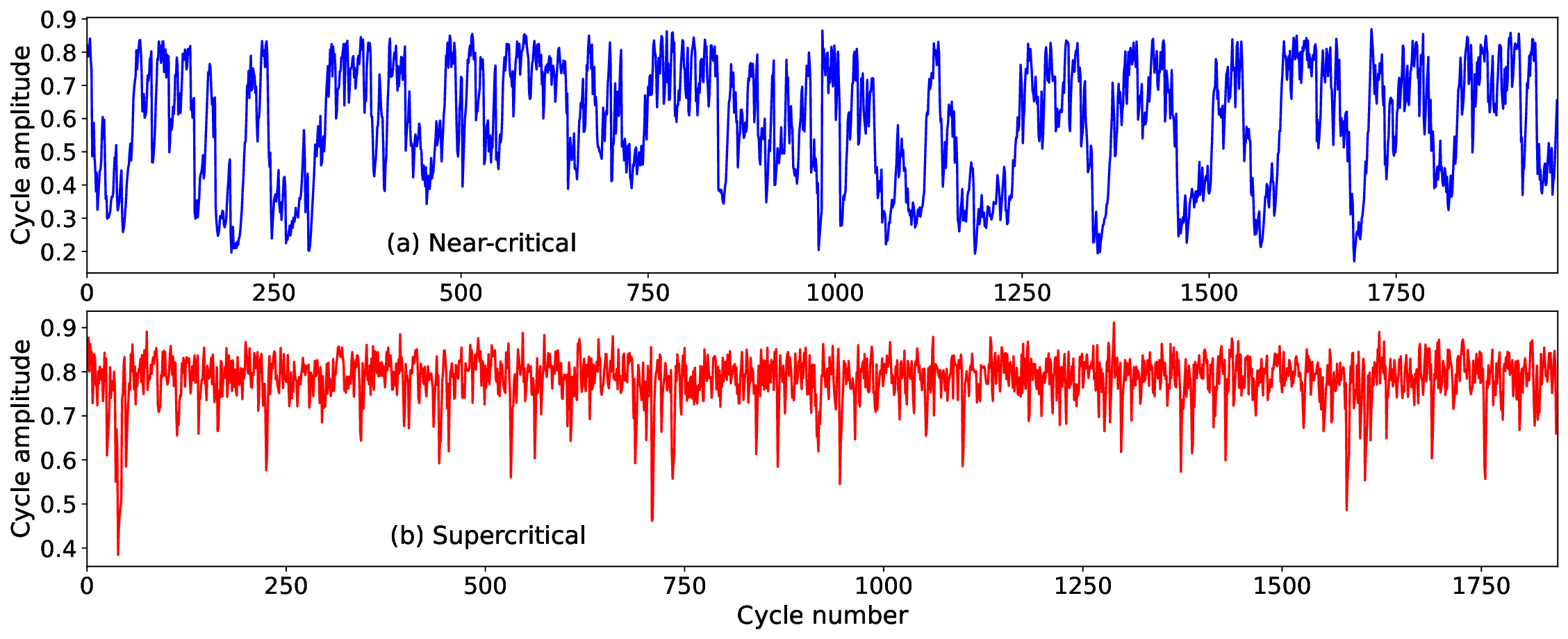}
    \caption{Time series of the peak values of the cycles of the toroidal flux obtained from Model I, operating in the (a) near-critical and (b) supercritical regimes.}
    \label{fig:ts_plot}
\end{figure*}

We have used the data from three \bl\ type dynamo models, namely Models I, II, and Time Delay dynamo, operating them in the near critical and supercritical regimes. Models I and II are essentially flux transport dynamo models built using the Surya code \citep{NC02, CNC04, C18} in which a local $\alpha$ prescription, single-cell meridional flow, turbulent diffusion, and differential rotation are used. The difference between Models I and II is that diffusion dominates in Model I (parameters
for the poloidal field diffusion: 
$\eta_2 = 1 \times 10^{12}$~cm$^2$~s$^{-1}$ and
$\eta_0 = 2 \times 10^{12}$~cm$^2$~s$^{-1}$
and for the meridional circulation:
$v_0 = 15$~m~s$^{-1}$), and advection dominates in Model II (same diffusivity parameters but $v_0 = 26$~m~s$^{-1}$).
\citet{YNM08} named these two models as diffusion and advection-dominated models, respectively.
To operate these models in near-critical and supercritical regimes, we take $\hat{\alpha}_0 = \alpha_0 / \alpha_0^{\rm crit} = 2$ and $4$, respectively (where $\alpha_0^{\rm crit}$ is the minimum $\alpha_0$ needed to obtain dynamo transition).
For the time delay dynamo model, we follow the one presented in \citet{wilsmith}.
The time delay model is also of \bl\ type in which the delays involved in communicating the toroidal and poloidal fields to their respective source regions are captured by suitable time delays in their sources in the differential equations \citep[Sec. 6.4 of][]{Kar23}. 
The details of all the models 
are presented in \citet{Pawan21b}.
For operating the time delay dynamo model in near critical and supercritical, we took $\hat{\alpha}_0 = 1$ and $3$, 
respectively. 
We note that the time delay model quickly goes to supercritical regime \citep[e.g., see Figure~5 of][]{Pawan21b}  and thus, to operate the model in near-critical and supercritical regimes, we have taken the value of $\hat{\alpha}_0$ as 1 and 3 instead of 2 and 4 as taken in models I and II.
From each model, we first compute the absolute value of the toroidal flux at low latitude at the base of the convection zone as a function of time
 (toroidal field in case of time delay model). 
Then, the peak values of the cycles of the toroidal flux/field are taken as the time series for our analyses.  
A representative example of the time series for Model~I at critical regime is shown in \Fig{fig:ts_plot}. 
The number of data points in each time series lies between 1800 and 2800.

Let us comment on the solar cycles from our models I and II. So far, no dynamo model is completely realistic \citep{Char20, Kar23}; our present ones are not the exception. Our dynamo models are axisymmetric and kinematic. The magnetic buoyancy and the Babcock-Leighton process are parameterised in simplified manners \citep{NC02, Biswas22prl}. However, the models still reproduce some basic features of the solar cycles, namely, the regular polarity reversals, a strong correlation between the polar field at the minimum and the amplitude of the next cycle toroidal field, a strong correlation between the rise rate and the amplitude of the cycle (Waldmeier effect), long-term modulation of the amplitude, grand minima and maxima. To highlight some of these features, in \Figs{fig:app1_cr}{fig:app1_scr} of Appendix, 
we show the variation of the toroidal field and the butterfly diagram from Model I at near-critical and supercritical regimes. We can already see from these figures that the long-term modulation of the cycle in the two regimes is different. The near-critical model produces a strong long-term modulation and frequent grand minima-like events. Our aim is to quantify this change of solar cycle memory more rigorously and to identify at what parameter regime of dynamo supercriticality, the model reproduces the observation best. 

\section{Methods} \label{sec:methods}
Nonlinear time series analysis techniques have been employed in various diverse fields, starting from the prediction in the stock market to understand the dynamics of various complex systems \citep{C13, Sanz22}. 
In astronomical and solar data, these have been used to identify the existence of low-dimensional chaotic or stochastic signature in the underlying processes \citep{M91,1994A&A...290..983C, ZQ98, H10, KarakDuttaMukhopadhyay10}, and the persistence of the memory in the solar cycle \citep{Maddanu2022}.  Recently, \citet{Das22} identified some memory in the solar cycle asymmetry data using different nonlinear dynamics parameters like correlation dimension and fractal dimension techniques. 
In the present work, we will exploit the nonlinear time series analysis to investigate the memory of the solar cycles in the dynamo models.

\subsection{Higuchi's dimension}


Fractal dimension $D$, used to characterize nonlinear time series, is computed using the method given in \citet{HIGUCHI1988277}. Here we only briefly discuss it. 

We start with a time series $X(i)$ containing $N$ observations that have been sampled at regular intervals. Thus,
\begin{equation} 
X(i): X(1), X(2), X(3), ..., X(N).
\end{equation}

From $X(i)$, we construct a new time series
\begin{equation} 
X^m_s: X(m), X(m+s), X(m+2s), ..., X\left(m+ns\right),
\end{equation}
where, $ 1 \le m \le s$ and $n = \left[\frac{N-m}{s}\right]$ denotes the greatest integer less than or equal to $\frac{N-m}{s}$.
Associated with each $X^m_s$, we calculate the length of the curve $L_m(s)$ as

\begin{eqnarray}
L_{m}(s) = 
\left\{\left(\sum_{i=1}^{n}|X(m+is)- X(m+(i-1)s)|\right) k \right\}\frac{1}{s}, \nonumber \\
\end{eqnarray}
where 
$k=(N-1)/ns$.

The average length of the curve $\langle L(s)\rangle$ is obtained by taking the mean of $L_{m}(s)$ for $1 \le m \le s$. For $s_{min} < s < s_{max}$, if we obtain $\langle L(s) \rangle \propto s^{-D}$, the time series is a fractal of dimension $D$ in that range of $s$. 
We compute $\langle L(s) \rangle$ for $2 \le s \le 32$ and $2 \le s \le 256$ for the data obtained from critical and supercritical dynamo regimes, respectively. 
$D$ is computed from the slope of the double logarithmic plot of $\langle L(s) \rangle$ vs $s$. 

The value of $D$ is a fraction with $1 < D < 2$. A value close to 2 denotes a space-filling curve, while a value close to 1 is a straight line.

\subsection{Hurst Exponent}


We follow the $R/S$ method introduced by \citet{Mandelbrot1969SomeLP} to find the Hurst exponent $H$.

We have a time series, $X(i), i = 1, 2, ..., N$, whose Hurst exponent we want to compute. Now, choose a temporal window $s$, with $s_t < s < N$. Here, $s_t$ is the Theiler window.
Now, we make $(N-s+1)$ subsets of the series $X(i)$ as follows:

\begin{equation} 
x_{t_0}(s): X(t_0), X(t_0+1), X(t_0+2), ..., X(t_0+s-1),
\end{equation}

where, $t_0 = 1, 2, ..., N-s+1$. Now, average of the subset $x_{t_0}(s)$ is given by

\begin{equation} 
\overline{x}_{t_0}(s)=\frac{1}{s}\sum_{i=t_0}^{t_0+s-1} X_i.
\end{equation}

Now, the standard deviation of $x_{t_0}(s)$ corresponding to window $s$ is given by
\begin{equation} 
S(t_0, s)=\sqrt{\frac{1}{s-1}\sum_{i=t_0}^{t_0+s-1}\left[X(i)-\overline{x}_{t_0}(s)\right]^2}.
\end{equation}

We define the set of cumulative deviations of $x_{t_0}(s)$ from the mean as
\begin{equation} 
y_i(t_0,s)=\sum_{k=t_0}^{t_0+i-1}\left[X(k)-\overline{x}_{t_0}(s)\right],
\end{equation}
where, $i = 1, 2, ..., s$. Thus, the Range of $y_i(t_0,s)$ is
\begin{equation} 
R(t_0,s)=\max_{1\leq i\leq s}y_i(t_0,s) - \min_{1\leq i\leq s}y_i(t_0,s).
\end{equation}
and the rescaled range measure $R/S$ is given by
\begin{equation} 
(R/S)(t_0,s)=\frac{R(t_0,s)}{S(t_0,s)}.
\end{equation}
For a particular $s$, we have $1 \le t_0 \le N-s+1$ and thus, the corresponding rescaled range is
\begin{equation} 
(R/S)_s=\frac{1}{N-s+1}\sum_{t_0}(R/S)(t_0,s).
\end{equation}
This value of the rescaled range is found to vary as
\begin{equation}
(R/S)_s=ks^H,
\end{equation}
where, $k$ is a constant and $H$ is the Hurst exponent. A plot is made for $\log(R/S)$ vs $\log(s)$ for $1 \le s \le N$ and the linear portion of the graph is fitted to obtain the Hurst exponent. 
A time series obtained from a white noise process yields $H = 0.5$. When $H > 0.5$ for a time series, the series is said to be persistent or has a long-term memory.
An increase in the value of the time series at a particular step is more probable to be followed by its increase in the next step (rather than a decrease). 
When $H < 0.5$ for a time series, it is said to be anti-persistent or has a short-term memory. 
An increase in the value of the time series at a particular step is more probable to be followed by its decrease in the next step.

\subsection{Multifractal Analysis}



Many nonlinear systems, including solar activity, are characterized by intermittent phenomena and in such scenarios, a single Hurst exponent is not sufficient to capture the essential characteristics of the system. The Multifractal Detrended Fluctuation Analysis (MF-DFA) helps reveal the complexity or multifractal structure of the time series by characterizing the amplitude fluctuations in the data. 
The MF-DFA method has been developed from Detrended Fluctuation Analysis (DFA) by \citet{KANTELHARDT200287}. This method has been applied to understand the underlying dynamics that lead to the flux variability of quasars \citep{belete18} and the nature of solar flare activity \citep{McAteer_2007, Sen2007-ad}.

We have a time series, $X(i), i = 1, 2, ..., N$, and we want to understand its multifractal structure. Let $\langle X \rangle$ be the mean of the time series $X(i)$. We compute the profile as
\begin{equation}
    Y(i) = \sum_{j=1}^{i} (X(i) - \langle X \rangle), i = 1, 2, ..., N
\end{equation}
Next, we choose a segment size $s$ and divide the new series $Y(i)$ into $N_s = [N/s]$ segments, where $[k]$ denotes the integer less than or equal to any real number $k$. Starting from the first element of the time series $Y(i)$, we obtain $N_s$ segments, each of size $s$ and will be left with $N$ (mod $s$) 
elements at the last, which do not belong to any segment. To account for the remaining part of the series, we repeat the same process by starting from the end of the time series, obtaining $N_s$ segments, and leaving $N$ (mod $s$) elements at the beginning. Hence, in total, we obtain $2 N_s$ segments. For each of these $2N_s$ segments of length $s$, we compute the first-order polynomial fit $y_\nu (i), 1 \le i \le s$ using the least square method. Next, the variance for each of the segments is computed as
\begin{equation}
    F^2(\nu, s) = \frac{1}{s} \sum_{i=1}^{s} \{ Y[(\nu-1)s+i] - y_{\nu}(i)\}^2 
\end{equation}
for $\nu = 1, 2, ..., N_s$, and as
\begin{equation}
    F^2(\nu, s) = \frac{1}{s} \sum_{i=1}^{s} \{ Y[N-(\nu-N_s)s+i] - y_{\nu}(i)\}^2 
\end{equation}
for $\nu = N_s+1, N_s+2, ..., 2N_s$.

We now compute the average variance of each of the segments as
\begin{equation}
    F_q(s) = \left \{ \frac{1}{2N_s} \sum_{\nu = 1}^{2N_s} [F^2(\nu, s)]^{q/2} \right \}^{1/q}, q \neq 0
\end{equation}
\begin{equation}
    F_q(s) = \exp \left \{ \frac{1}{4N_s} \sum_{\nu = 1}^{2N_s} \ln [F^2(\nu, s)] \right \}, q = 0
\end{equation}

We are interested in knowing how the average variance $F_q(s)$ scales with the size of the segments $s$ for different values of $q$. Suppose
\begin{equation}
    F_q(s) \sim s^{h(q)}
    \label{eqn:avg_fluctuation_scaling}
\end{equation}
and we are interested to find the value of the exponent $h(q)$ known as the generalized Hurst exponent. 
For very small segment size $s \approx 10$, there are systematic deviations from the scaling behavior, while for large segment size $s > N/4$, the method is statistically unreliable \citep{KANTELHARDT200287, Ihlen2012}. Thus, we have chosen the segment size $s$, such that $15 \le s \le N/4$.
For $q = 2$, the method corresponds to the method of Detrended Fluctuation analysis (DFA) and the generalized Hurst exponent $h(q = 2)$ corresponds to the Hurst exponent $H$.

For any monofractal time series, the generalized Hurst exponent is independent of $q$ (or varies weakly with $q$). While a strong $q$ dependence of $h(q)$, indicates the time series is multifractal. Moreover, if a homogeneous scaling behavior of $F_q(s)$ (in \Eq{eqn:avg_fluctuation_scaling}) is obtained, the $h(q)$ values for $q < 0$ are usually larger than the values when $q > 0$.

This generalized Hurst exponent $h(q)$ is related to the scaling exponent $\tau(q)$ of the standard partition function based multifractal formalism, by
\begin{equation}
    \tau(q) = q h(q) - 1
\end{equation}
Difference between the slope of $\tau(q)$ at $q < 0$ and $q > 0$ indicates the strength of multifractality in the data. 

Another way to see the multifractal structure of the data is to use the singularity spectrum given by 
\begin{equation}
    f(\alpha) = q\alpha - \tau(q)
\end{equation}
where, $\alpha = \tau' = \frac{d\tau}{dq}$. Here, $\alpha$ denotes the singularity strength and $f(\alpha)$ denotes the dimension of the subset of the time series that is characterized by $\alpha$.

Once we have the multifractal spectrum, we can note down the minimum (maximum) value of the singularity strength $\alpha_{\textrm{min}} (\alpha_{\textrm{max}})$. The \textit{width} of the multifractal spectrum is given by $\Delta \alpha = \alpha_{\textrm{max}} - \alpha_{\textrm{min}}$. A broad spectrum, indicated by large values of $\Delta \alpha$, represents a higher degree of multifractality in the data \citep{Ihlen2012}. In the monofractal limit, the spectrum reduces to a single point and $\Delta \alpha$ goes to zero. However, for real-world signals which always have a finite length, the multifractal spectrum always has a small (non-zero) width.

The value of $\alpha$ at which $f(\alpha)$ assumes a maximum value is denoted by $\alpha_0$. The symmetry in the shape of the spectrum is given by the quantity $A = (\alpha_{\textrm{max}} - \alpha_0)/(\alpha_0 - \alpha_{\textrm{min}})$. For a right-skewed (left-skewed) spectrum, we have $A > 1$ $(A < 1)$. When $A = 1$, the multifractal spectrum is symmetric. A right-skewed (left-skewed) multifractal spectrum arises due to left (right) truncation. A left (right) truncation of the multifractal spectrum occurs due to the leveling of the $q^{\textrm{th}}$-order Hurst exponents for $q > 0$ $(q < 0)$ \citep{Ihlen2012}. A left (right) truncation indicates that the multifractal structure is sensitive to small-scale temporal fluctuations with small (large) amplitudes. Moreover, a left (right) truncation shows a higher abundance of small (large) amplitude fluctuations in the time series. 

\section{Results and Discussion}
\label{sec:results}
\subsection{From model data}
\begin{figure}
\centering
\includegraphics[width=\columnwidth]{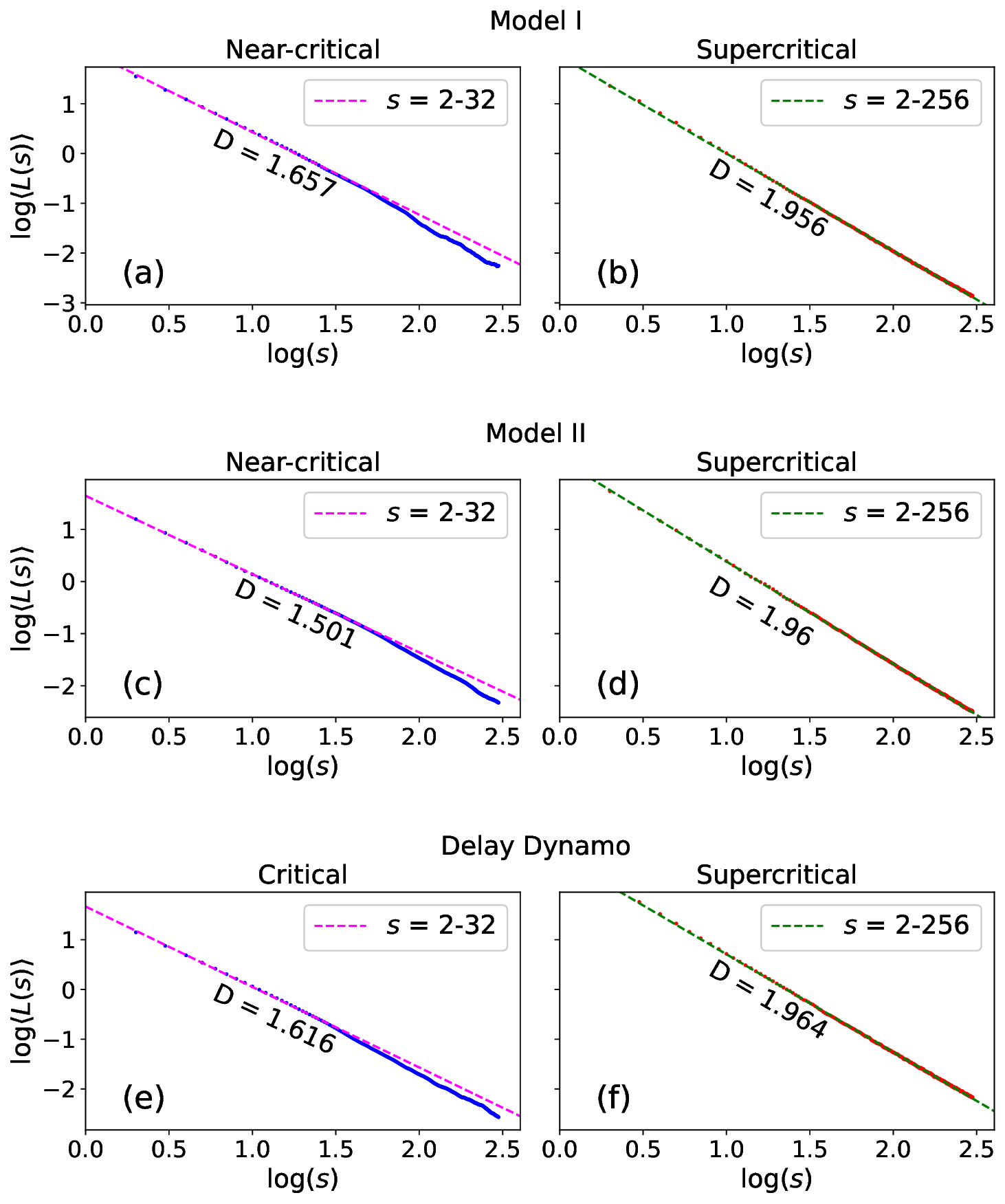}

\caption{Variation of length $\langle L(s)\rangle$ with time interval ($s$) for toroidal flux data obtained from dynamo models in critical (left) and supercritical (right) regimes.
}
\label{fig:higuchi}
\end{figure}


To explore how the nature of the magnetic cycle changes with the supercriticality of the solar dynamo, we shall now look into the results from the nonlinear time series analysis methods as discussed above. 

For Model I, we find $D = 1.657$ and $D = 1.956$ for the time series in the near-critical (\Fig{fig:higuchi}a) and supercritical (\Fig{fig:higuchi}b) regimes, respectively. Again, for Model II, the time series in the near-critical and supercritical regimes produce $D = 1.501$ (\Fig{fig:higuchi}c) and $D = 1.972$ (\Fig{fig:higuchi}d), respectively. Finally, for the Delay Dynamo Model, we obtain $D = 1.616$ and $D = 1.964$ when the dynamo operates in the critical (\Fig{fig:higuchi}e) and supercritical (\Fig{fig:higuchi}f) regime, respectively.

A periodic time series will be fractal dimension $D = 1$, and a highly stochastic time series will be fractal dimension $D = 2$ \citep{HB18}. In the critical regime, a value of the fractal dimension $D$ close to 1.5 shows that the system may be chaotic and irregular. However, in the supercritical regime of the dynamo, the fractal dimension is close to 2, and thus, the process must be stochastic.

Moreover, we note that when the dynamo operates near the critical regime, the value of the fractal dimension is obtained at a timescales of about 2-32 solar cycles. As we look into larger timescales (beyond 32 solar cycles), the slope changes, which indicates that some different fractal dimension is needed to characterize the time series at these larger timescales. These are typical of structures that are not self-similar. 
However, when the dynamo operates in the supercritical regime, a single fractal dimension is obtained in a wide range of timescales, ranging from 2 to 256 solar cycles. This is indicative of the self-similar nature of the time series obtained in the supercritical regime but not in the critical regime.

\begin{figure}
    \centering
    \includegraphics[width=\columnwidth]{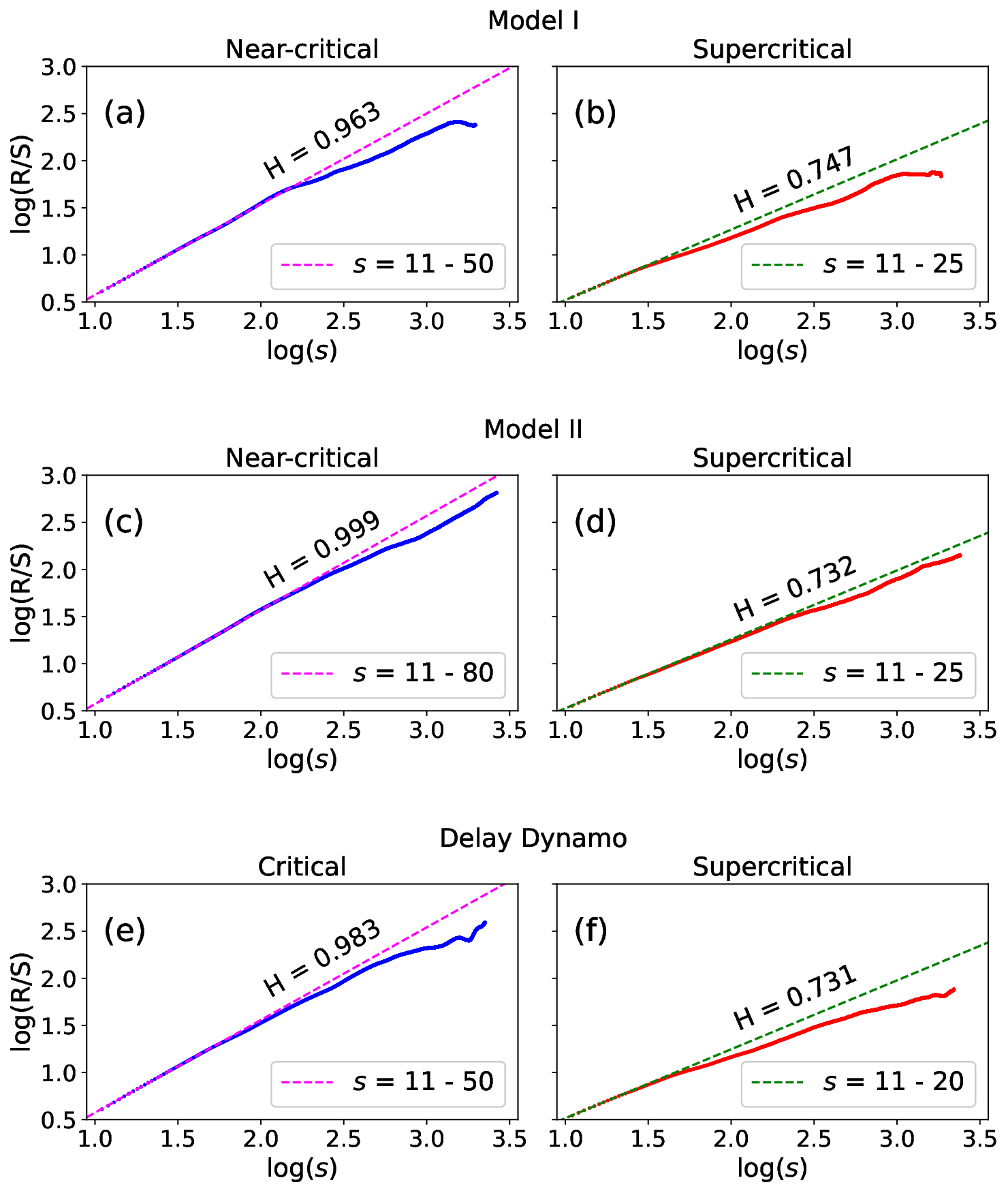}
    \caption{Variation of $R/S$ with time interval ($s$) for toroidal flux data obtained from different models in the near-critical (left) and supercritical (right) regimes.}
    \label{fig:hurst}
\end{figure}

A time series having a fractal dimension between $1.0$ to $2.0$ will have a memory effect \citep{SZ17}. The range of fractal dimensions for the time series are varying from $1.50$ to $1.97$ as we go from near-critical to supercritical regimes of the dynamo. 
It gives a clue about the persistence of memory. Finally, we computed the Hurst exponent to look for the persistence in the solar cycle memory.

For Model I, we see that the Hurst exponent in the near-critical regime is $0.963$ for the range $s = 11 - 50$ and in the supercritical regime is $0.747$ for the range $s = 11 - 25$ (\Fig{fig:hurst}a-b) which is $> 0.5$, i.e., irregular dynamo time series contains memory. We also see that in the near-critical regime, the value of the Hurst exponent is greater than in the supercritical regime. Moreover, the number of solar cycles over which the memory persists is greater when the dynamo operates in the near-critical regime (11--50) compared to when it operates in the supercritical regime (11--25). 

For Model II, the Hurst exponent in the near-critical regime is $0.999$ for the range $s = 11 - 80$ and in the supercritical regime is $0.745$ for the range $s = 11 - 25$ (\Fig{fig:hurst}c-d). 
For the time delay dynamo model, the Hurst exponent in the critical regime is $0.983$ for the range $s = 11 - 50$ and in the supercritical regime is $0.731$ for the range $s = 11 - 20$ (\Fig{fig:hurst}e-f).

Thus, all the dynamo models give similar results for the Hurst exponents. We observe that a dynamo operating in the near-critical regime has a stronger memory that persists over a higher number of solar cycles when compared to the memory in a dynamo operating in the supercritical regime. This supports our previous conclusion made based on the correlations between the poloidal and toroidal fields \citep{Pawan21b}.

\begin{figure}
    \centering
    \includegraphics[width=\columnwidth]{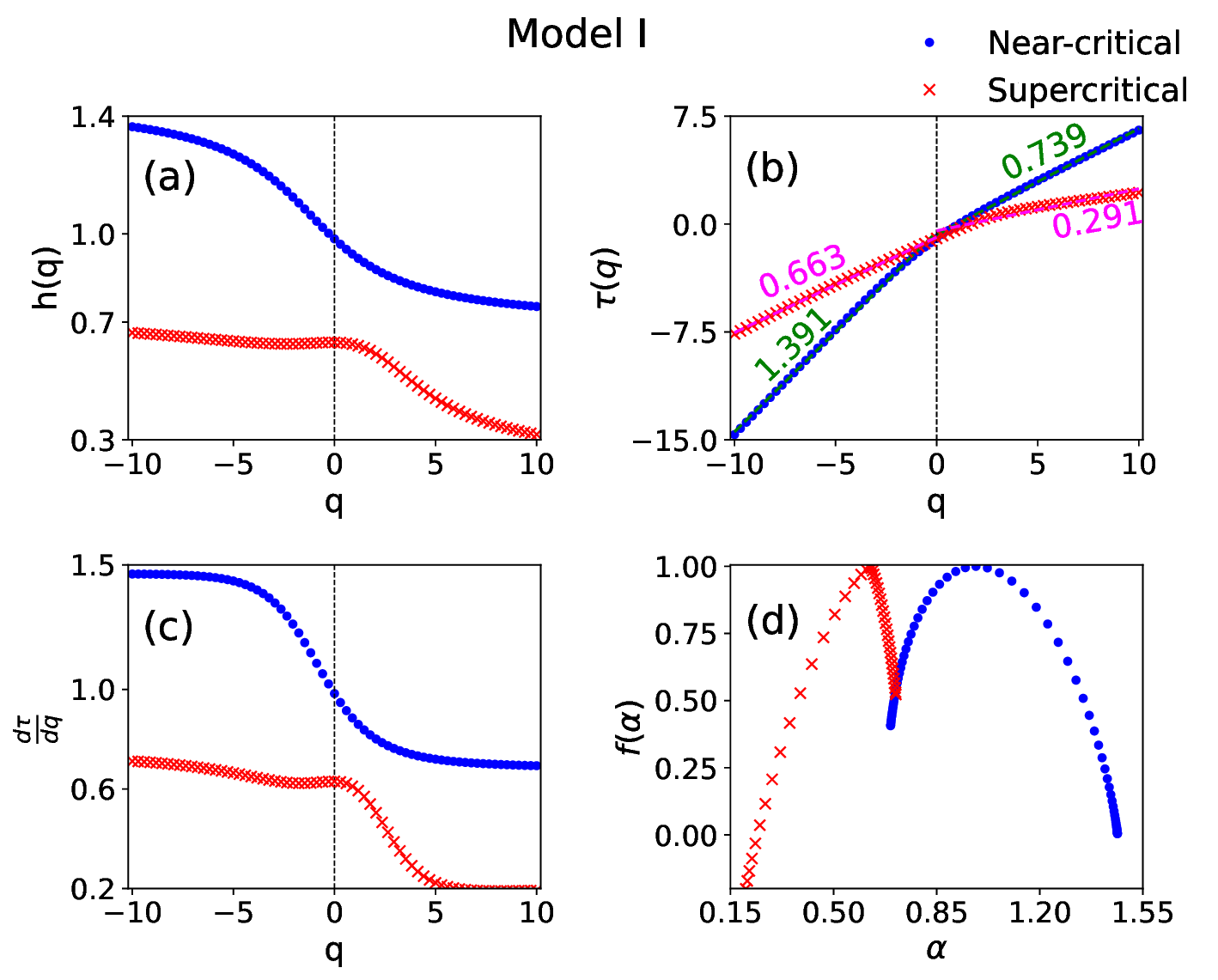}
    \caption{Results from the MF-DFA method obtained in the near-critical and supercritical regimes of Model I.}
    \label{fig:mfdfa_modelI}
\end{figure}

The results from the MF-DFA method for Model I are shown in \Fig{fig:mfdfa_modelI}. The generalized Hurst exponent depends on the value of $q$, which shows that both the time series have some degree of multifractality. However, for the near-critical regime, the dependence of $h(q)$ on $q$ is stronger than in the super-critical case, implying the multifractal structure is pronounced in the near-critical regime. 
For $q=2$, the generalized Hurst exponent $h(q) = 0.890$ and 0.603 respectively in the near-critical and supercritical regimes. This shows a large persistence of the time series in the near-critical regime compared to the supercritical regime. The result is qualitatively similar to what has been obtained in the $R/S$ method. But, the values of the Hurst exponent may be slightly smaller since we are looking at much larger timescales in the MF-DFA method. The magnitude of the change in slope of the graph of $\tau(q)$ versus $q$ at $q = 0$ indicates the multifractal structure in the time series. The change in slope in the near-critical (0.652) and supercritical (0.372) regime can also be seen from the jump in the value of $\frac{d\tau}{dq}$ around $q = 0$ in \Fig{fig:mfdfa_modelI}c. Finally, we obtain the multifractal spectrum ($f(\alpha)$ vs $\alpha$) with left (right) truncation when the dynamo operates in the near-critical (supercritical) regime. The widths of the spectrum ($\Delta \alpha$) in the near-critical (0.770) and supercritical (0.521) regimes also tell us that the multifractality is higher in the near-critical regime when compared to the supercritical regime.

\begin{figure}
    \centering
    \includegraphics[width=\columnwidth]{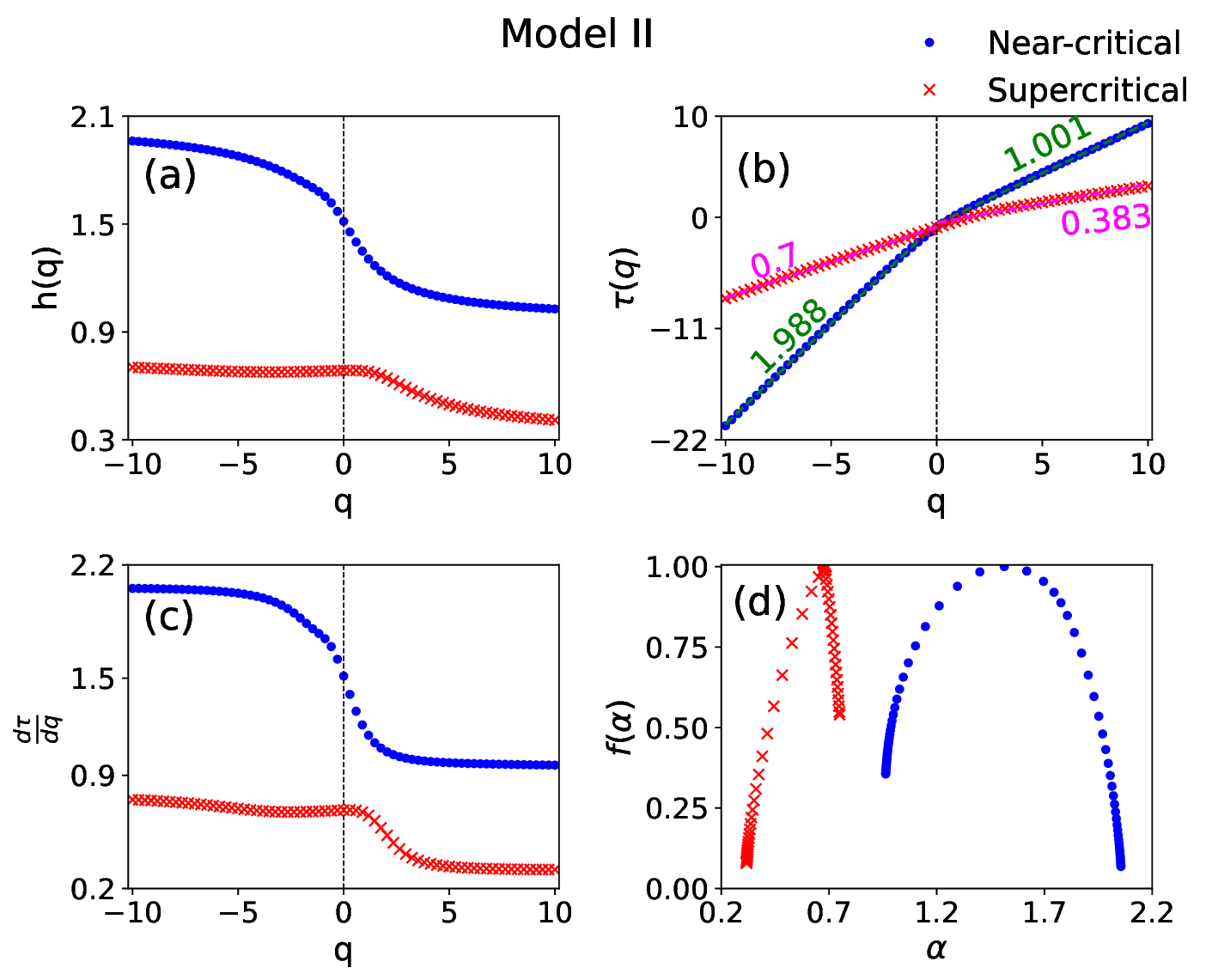}
    \caption{Same as \Fig{fig:mfdfa_modelI} but for Model II.}
    \label{fig:mfdfa_modelII}
\end{figure}

\Fig{fig:mfdfa_modelII} shows the results from the MF-DFA method when applied to the time series obtained from Model II. Again, we notice a stronger dependence of the generalized Hurst exponent $h(q)$ on $q$ in the near-critical regime. For $q=2$, the generalized Hurst exponent $h(q) = 1.24$ in the near-critical regime and $h(q) = 0.659$ in the supercritical regime. The value of $h(q=2) > 1$ implies that the non-stationarity in the data could not be successfully removed by detrending in the near-critical regime \citep{Bryce2012, ceballos2018estimation}. However, the value of $h(q)$ in the supercritical regime suggests little or no memory in this regime. 
The change in slope of $\tau(q)$ vs $q$ around $q = 0$ is higher in the near-critical regime (0.987) compared to the supercritical regime (0.317). The multifractal spectrum ($f(\alpha)$ vs $\alpha$) is again obtained with left (right) truncation when the dynamo operates in the near-critical (supercritical) regime. The width of the spectrum ($\Delta \alpha$) also tells us that the multifractality is higher in the near-critical regime (1.092) when compared to the supercritical regime (0.432).

\begin{figure}
    \centering
    \includegraphics[width=\columnwidth]{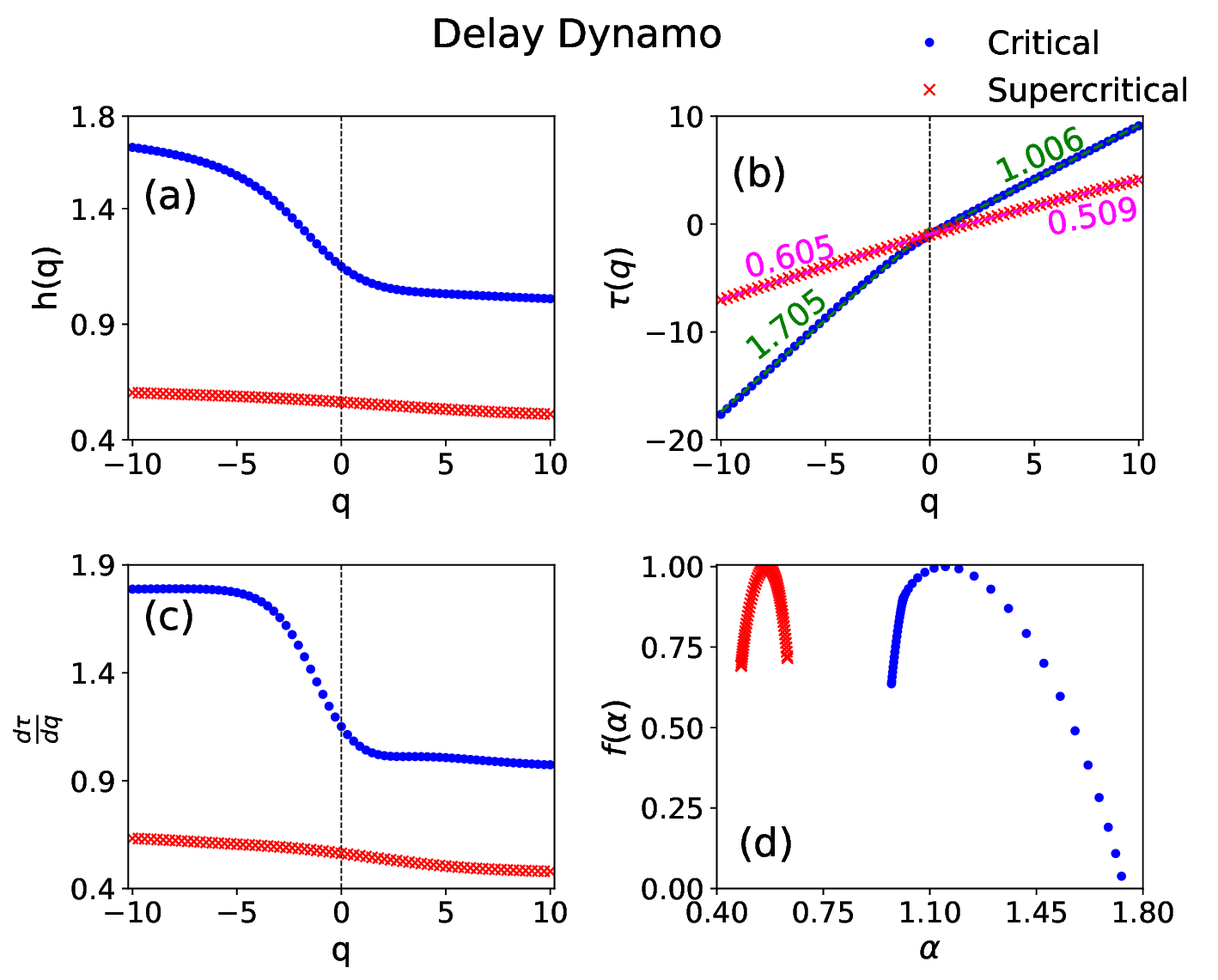}
    \caption{Same as \Fig{fig:mfdfa_modelI} but for the delay dynamo model.}
    \label{fig:mfdfa_delay}
\end{figure}

\Fig{fig:mfdfa_delay} shows the results of the MF-DFA method from the Delay Dynamo Model. As obtained earlier, the generalized Hurst exponent $h(q)$ strongly depends on $q$ in the near-critical regime, implying a higher degree of multifractality. We also find that $h(q = 2) = 1.07$ in the near-critical regime and $h(q = 2) = 0.552$ in the supercritical regime. 
Again, a value of $h(q=2)$ greater than unity suggests that the non-stationarity or trend in the data from the near-critical regime could not be successfully removed. However, a value slightly larger than unity may be interpreted as high persistence in the data, similar to that obtained by the $R/S$ method (0.983). In the supercritical regime, a value close to $0.5$ indicates little or no memory.
The slope of $\tau(q)$ vs $q$ changes more sharply around $q = 0$ in the near-critical regime (0.699) compared to the supercritical regime (0.096); 
also, see \Tab{tab:res_modeldata} for the key parameters of MF-DFA methods from all models for comparison.
The width of the spectrum ($\Delta \alpha$) confirms that the multifractality is higher in the near-critical regime (0.874) when compared to the supercritical regime (0.166). The multifractal spectrum is obtained with a left truncation in the near-critical regime, whereas it is mostly symmetric in the supercritical regime.

The unequal abundance of the small and large amplitude variations gives rise to the multifractal structures in the magnetic field data obtained from the solar dynamo models.
A higher multifractality in the near-critical regime is due to the relatively larger inequality in the abundance of small and large amplitude variations in this regime compared to the supercritical case.
Moreover, in the near-critical regime, the multifractal spectrum is obtained with left truncation, implying the multifractal nature of the time series arises due to the higher abundance of local fluctuations with small-scale amplitude variations.
In the supercritical regime, the width of multifractal spectrum is narrow and is obtained with occasional right truncation. This suggests that large amplitude fluctuations may be more common but not by a large margin compared to the small amplitude fluctuations. 
Thus, the time series in this regime exhibits close to monofractal behavior. 

\begin{table*}
    \centering
    \begin{tabular}{|c|c|cccc|}
    \cline{1-6}
         Dynamo regimes & Parameter  & Window (cycles) & Model I & Model II & Delay Dynamo \\
    \cline{1-6}
         Near-critical & $D$ & 2-32 & 1.657 & 1.501 & 1.616 \\
         & $H$ & 11-50/11-80 & 0.963 & 0.999 &  0.983 \\
         & $\Delta \alpha$ & 15-365 & 0.770 & 1.092 & 0.816 \\
         & Skewness ($A$) & 15-365 & Right (1.657) & Right (0.984) & Right (3.602) \\
    \cline{1-6}
         Supercritical & $D$ & 2-256 & 1.956 & 1.960 & 1.964 \\
         & $H$ & 11-20/11-25 & 0.747 & 0.732 &  0.731 \\
         & $\Delta \alpha$ & 15-365 & 0.521 & 0.432 & 0.152 \\
         & Skewness ($A$) & 15-365 & Left (0.205) & Left (0.214) & None (0.828) \\
    \cline{1-6}     
    \end{tabular}
    \caption{Summary of the results of Higuchi's dimension, Hurst exponent and the MF-DFA method for different model time series in near-critical and supercritical regimes.}
    \label{tab:res_modeldata}
\end{table*}

\begin{table}
    \centering
    \begin{tabular}{|c|ccc|}
    \cline{1-4}
        Parameter & Window (cycles) & $\langle {\rm SN} \rangle$ & Ensemble of $\langle {\rm SN} \rangle$ \\
    \cline{1-4}
        $D$ & 2-32 & 1.737 & 1.749 $\pm$ 0.015 \\
        $H$ & 11-22 & 0.857 & 0.843 $\pm$ 0.062 \\
        $\Delta \alpha$ & 11-21 & 3.010 & 2.609 $\pm$ 0.633 \\
        $A$ & 11-21 & 6.167 & 5.480 $\pm$ 1.989 \\
    \cline{1-4}
    \end{tabular}
    \caption{Summary of results obtained for Higuchi's dimension, Hurst exponent and the MF-DFA method from the reconstructed sunspot number.}
    \label{tab:res_solardata}
\end{table}

\subsection{From reconstructed sunspot number}

 Regular homogeneous data of sunspot number are available only for about the last 30 solar cycles \citep{Uso23}, 
which is too small to study the long-term behavior of the solar activity. 
However, recently, the annual solar activity series of the last millennium has been reconstructed from $^{14}$C data by \citet{usoskin_recons_solactivity_21}. 
They have also isolated the individual solar cycles and the 
corresponding cycle-averaged sunspot number, $\langle {\rm SN} \rangle$, during the period from 971 to 1899. This $\langle {\rm SN} \rangle$ is a measure of the cycle strength. 
The data was found to contain 85 solar cycles. 
We construct a time series of 
these 85 data of $\langle {\rm SN} \rangle$ and study the nonlinear characteristics of this time series. 
We note that as the sunspot data we use here is ``reconstructed'', which involves a sequence of model steps, its quality is compromised. However, as we are using the cycle-averaged SN in our analyses, the effect of noise is somewhat reduced. 
Furthermore, \cite{Uso23} showed that this reconstructed data closely resembles the available direct observations of solar cycles and reproduces the popular Waldmeier effect \citep{Wald, KC11}. Given these, we hope that this indirectly observed data carries some genuine features of the long-term behaviour of solar activity in the past, which can be used to compare with the results from dynamo models operating in two different regimes to get a hint of the operation of the solar dynamo.

\begin{figure}
    \centering
    \includegraphics[width=\columnwidth]{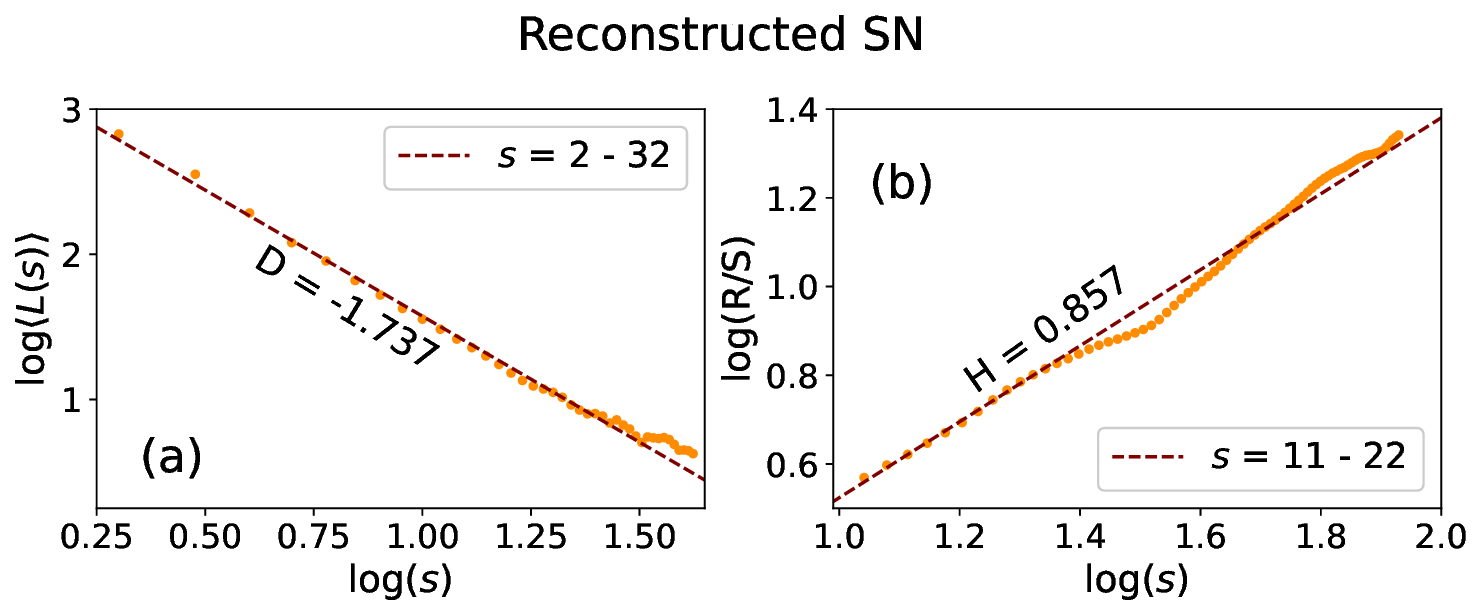}
    \caption{Variation of length $\langle L(s)\rangle$ (left) and $R/S$ (right) with time interval ($s$) for the reconstructed SN during the last millennium \citep{usoskin_recons_solactivity_21}.}
    \label{fig:D_H_solardata}
\end{figure}

\Tab{tab:res_solardata} tabulates
the results of Higuchi's dimension and Hurst exponent applied to the $\langle {\rm SN} \rangle$ time series. The Higuchi's dimension and Hurst $R/S$ for the $\langle {\rm SN} \rangle$ time series are obtained to be 1.737 and 0.857; see  \Fig{fig:D_H_solardata}. Due to the restricted size of the time series, we were limited to a smaller window range for the Hurst exponent compared to the results obtained from the Model data.
To account for the error in the $\langle {\rm SN} \rangle$ time series, we compute 100 resampled data sets with the same size as the $\langle {\rm SN} \rangle$ time series. An ensemble of 100 data points from a Gaussian distribution is produced, considering the $\langle {\rm SN} \rangle$ as the mean and the corresponding errors as the standard deviation. Finally, we compute the Higuchi's dimension and Hurst exponent for the 100 resampled time series and report their means and standard deviations in \Tab{tab:res_solardata}. The Higuchi's dimension and the Hurst exponent are found to be $1.749 \pm 0.015$ and $0.843 \pm 0.062$, respectively. This suggests that the data is not completely stochastic but may be irregular and chaotic. In addition, there is some memory in the data. 
However, the results do not precisely locate the mode of operation of the solar dynamo-- it only implies that the solar dynamo is operating somewhere between the near-critical and highly supercritical.

\begin{figure}
    \centering
    \includegraphics[width=\columnwidth]{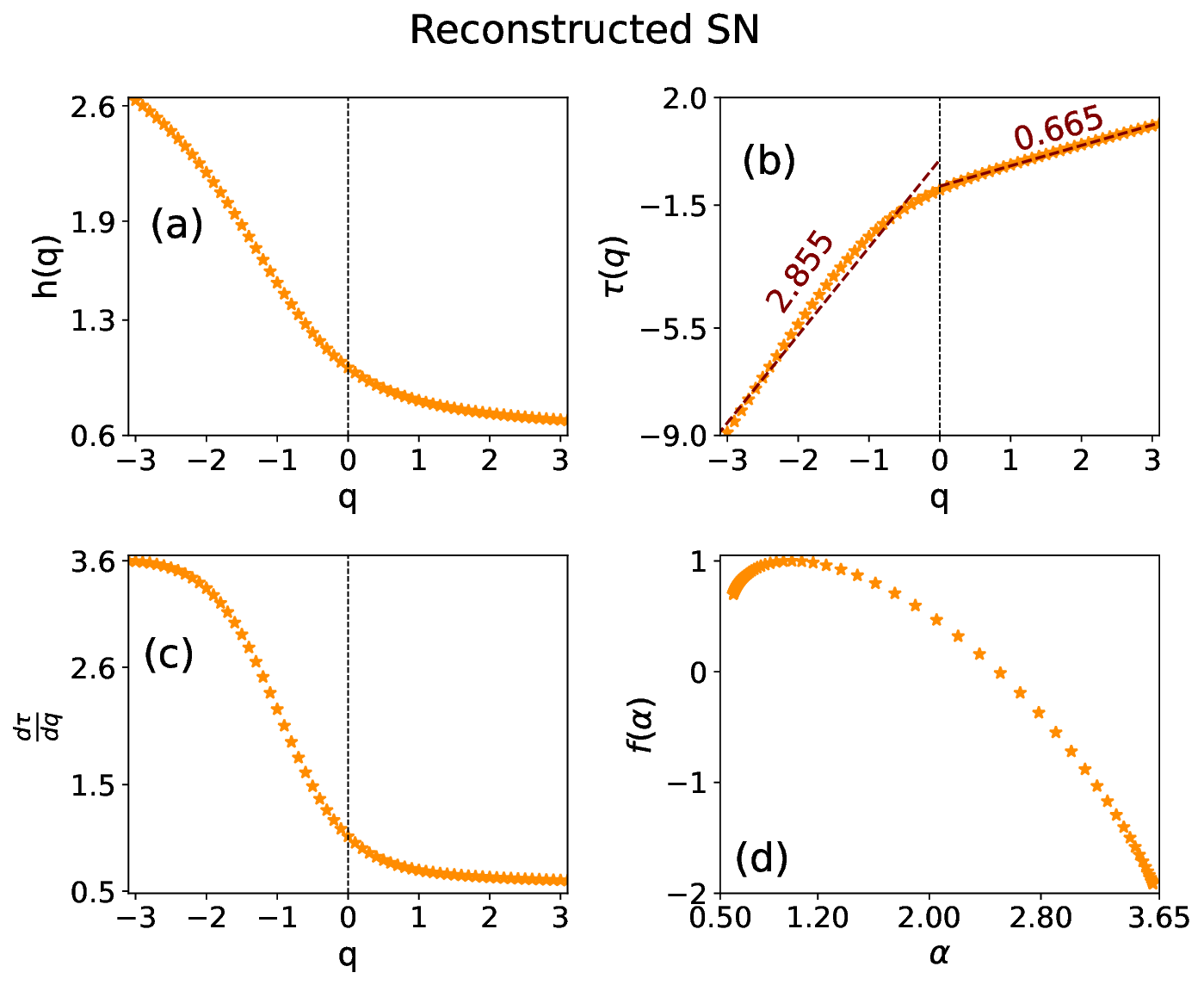}
    \caption{Results from the MF-DFA method obtained from the reconstructed SN.
    }
    \label{fig:mfdfa_solardata}
\end{figure}

\Fig{fig:mfdfa_solardata} shows the multifractal characteristic of the 
$\langle {\rm SN} \rangle$ time series. The time window (segment size $s$ in Section \ref{sec:methods}) is chosen to be 11-21 cycles. Though systematic deviations may arise at small time windows (11-14 cycles) in the MF-DFA method, we checked there were no qualitative differences in the results when we chose the time window to be 15-21 cycles. The generalized Hurst exponent $h(q)$ strongly depends on $q$ and changes from $2.6$ to $0.6$ as $q$ varies from $-3$ to $3$. The variation is mostly when $q < 0$. This shows a very high degree of multifractality in the data. At $q = 2$, the Hurst exponent is $0.734$, indicating there is some persistence in the time series. The slope of the $\tau(q)$ vs $q$ graph changes around $q = 0$ by a large amount (2.19). This change in the slope is also seen in \Fig{fig:mfdfa_solardata}c. The multifractal spectrum ($f(\alpha)$ vs $\alpha$) is obtained with a left truncation and the width of the spectrum ($\Delta \alpha$) is quite large (3.010). The spectrum is also found with a left truncation ($A = 6.167$). We repeat the MF-DFA analysis on the 100 resampled data sets. The spectrum does not follow the expected inverted parabolic shape for 13 data sets and is found to produce $f(\alpha) > 1$. This may arise when the homogeneous scaling, as assumed in \Eq{eqn:avg_fluctuation_scaling}, does not hold in the considered time window. From the remaining data sets, we find that the width of the multifractal spectrum is $\Delta \alpha = 2.609 \pm 0.633$ and the asymmetry parameter $A = 5.480 \pm 1.989$.

The large width of the spectrum suggests the need for a large number of exponents to characterize the data and, thus, the breakdown of self-similarity. The left truncation of the multifractal spectrum suggests the abundance of local fluctuations with small-scale amplitude variations. We can also say that the multifractal structure of the data is sensitive to small-scale temporal variations with small-amplitude. The sensitivity of the multifractal structure to the local fluctuations with small amplitude variations is also seen for the time series obtained from the dynamo models when they operate near the near-critical regime (left truncation of the $f(\alpha)$ vs $\alpha$ curve in Figures \ref{fig:mfdfa_modelI}d, \ref{fig:mfdfa_modelII}d and \ref{fig:mfdfa_delay}d). 
This suggests that the Sun operates close to the near-critical regime and not in the highly supercritical one. 

\section{Discussion and Conclusion} 
      \label{S-Conclusion} 

Using nonlinear time series analysis techniques, we have studied the behaviour of long-term variation of the magnetic cycles in near-critical and supercritical regimes of the solar dynamo. For this, we have considered the peak values of the toroidal flux from three different dynamo models. 
We find that Higuchi's fractal dimension ($D$) is close to 1.5 and 2, respectively, for the near-critical and supercritical regimes of the dynamo. 
On the other hand, the Hurst exponent ($H$) 
is near 1 and 0.74 for these two regimes. 
Values of $D$ and $H$ suggest that the magnetic cycle in the supercritical regime is governed by less memory and a more stochastic process. In other words, when dynamo operates in the supercritical regime, the persistence nature of the cycle is weak, and the stochastic nature is dominant.  
By observing the long window in $H$ (or in the scaling relation of $R/S$ vs $\tau$ as seen in \Fig{fig:hurst} and \Tab{tab:res_modeldata}), we conclude that the memory of the magnetic cycle in the near-critical regime is very long.  
This result is congruous with the previous expectation based on the linear correlation between the peaks of the polar field and that of the subsequent cycles \citep{Pawan21b}. 
Moreover, we see evidence of self-similarity in the time series obtained in the supercritical regime. However, this self-similarity breaks down when we approach the near-critical regime, as we see a single value of the fractal dimension is not good enough for a range of timescales. 


The Multi-Fractal Detrended Fluctuation Analysis (MF-DFA) tells us that the breakdown of self-similarity (or intermittency) in the near-critical regime arises due to the large abundance of small-amplitude fluctuations in the time series, with occasional large-amplitude variations. 
However, the time series in the supercritical regime has little or no disparity in the abundance of small and large amplitude variations. This gives rise to a weak multifractal or close to monofractal behavior in the supercritical regime.


We can understand the above results from the dynamo model in the following way. In the near-critical regime, the effect of nonlinearity is weak, and the dynamo number and, thus, the growth rate is small. In this regime, the dynamo takes a long time to grow the magnetic cycle if it falls to a low value. This produces long-term modulation and intermittent behaviour of the magnetic cycle. On the other hand, in the supercritical regime, the dynamo number and, thus, the growth rate and the nonlinearity are high. These cause the magnetic field to grow rapidly (due to the high growth rate) when it has fallen to a low value or decrease rapidly (due to strong nonlinearity) when the field has enhanced to a high value. These  break the long-term modulation, making the cycle more stochastic and tends to give a self-similar pattern.

Further, we compute the nature of the average solar cycle data obtained from annual solar activity series from the last millennium that has been reconstructed from $^{14}$C data \citep{usoskin_recons_solactivity_21}. 
 Given the considerable uncertainty in reconstructed data and the limited number of cycles, the computed values of $D$ and $H$ suggest that the solar dynamo is possibly not operating in a highly supercritical regime.  
The multifractal analyses show a lack of self-similarity and a high multifractal nature in the time series, arising due to the abundance of local fluctuations with small amplitude variation. These suggest that the solar activity is due to an underlying dynamo process that operates in the near-critical regime or, at least not, in the highly supercritical one.  This conclusion is in agreement with the previous independent investigations \citep{R84, Met16, KN17, V23}.  

\begin{acknowledgments}
The authors are indebted to the referee for carefully reviewing the manuscript and providing critical comments that helped improve the quality of the presentation. 
 AG acknowledges Kishore Vaigyanik Protsahan Yojana (KVPY) for scholarship during a part of the project. B.B.K. acknowledges the ﬁnancial support from the Department of Science and Technology (SERB/DST), India, through the Ramanujan Fellowship (project No. SB/S2/RJN-017/2018). 
\end{acknowledgments}

%

\vspace{5mm}








\bibliography{paper}{}
\bibliographystyle{aasjournal}

\begin{appendix}

\section{Additional information for the model solar cycles}
\label{sec:App}
\begin{figure}
    \centering
    \includegraphics[width=\textwidth]{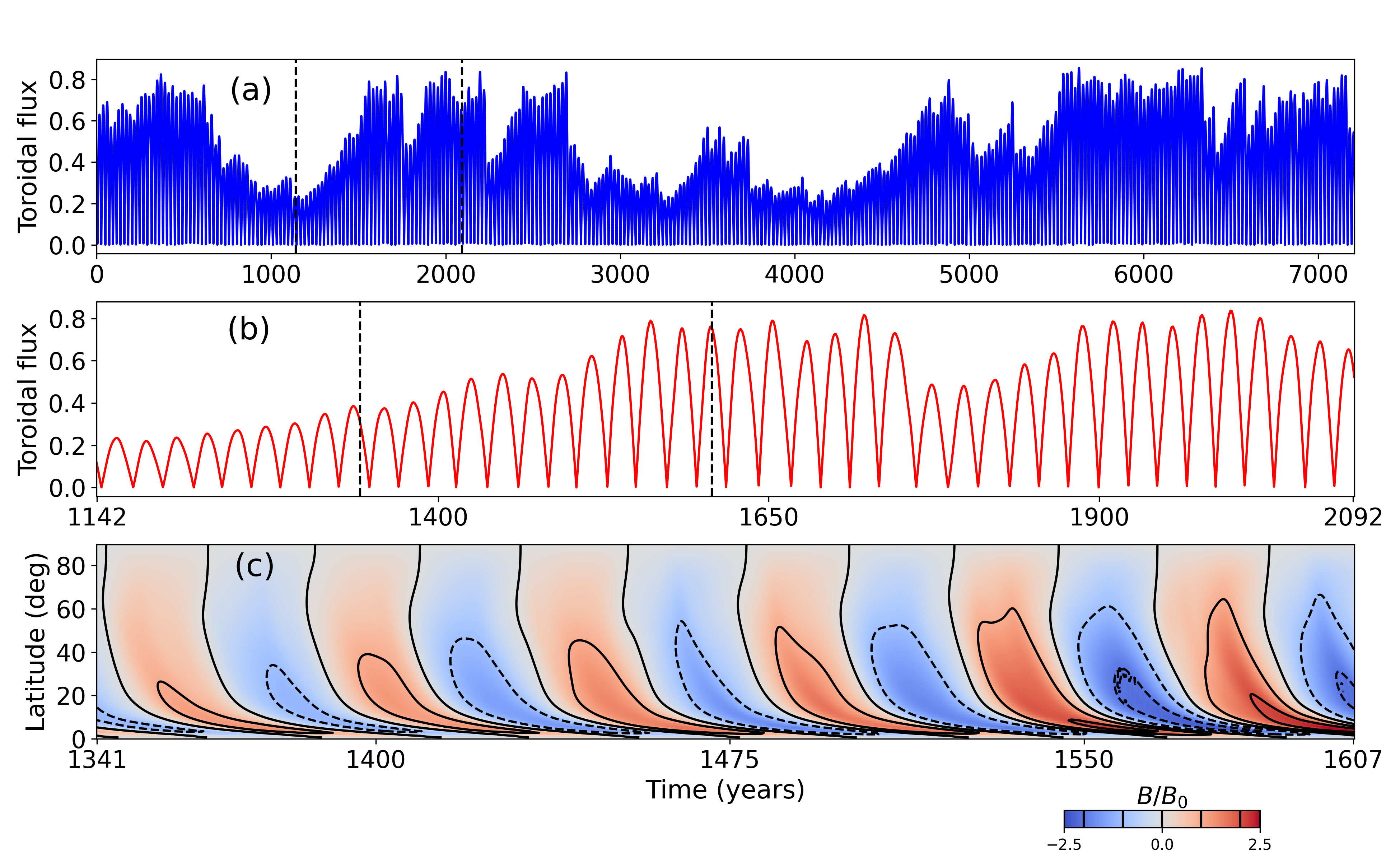}
    \caption{
    Results from Model I operating at near-critical regime.  (a) The temporal variation of the toroidal flux (in code unit) obtained within a radial extent of $0.67R_\odot$ to $0.72R_\odot$ and latitude extent of  $15^\circ$ to $45^\circ$.  (b) Highlighting the cycles marked by two vertical black dashed lines in (a).  (c) The time-latitude distribution (butterfly diagram) of the toroidal field at $0.72R_\odot$ during the time interval marked in (b) by vertical black lines.
    }
    \label{fig:app1_cr}
\end{figure}

\begin{figure}
    \centering
    \includegraphics[width=\textwidth]{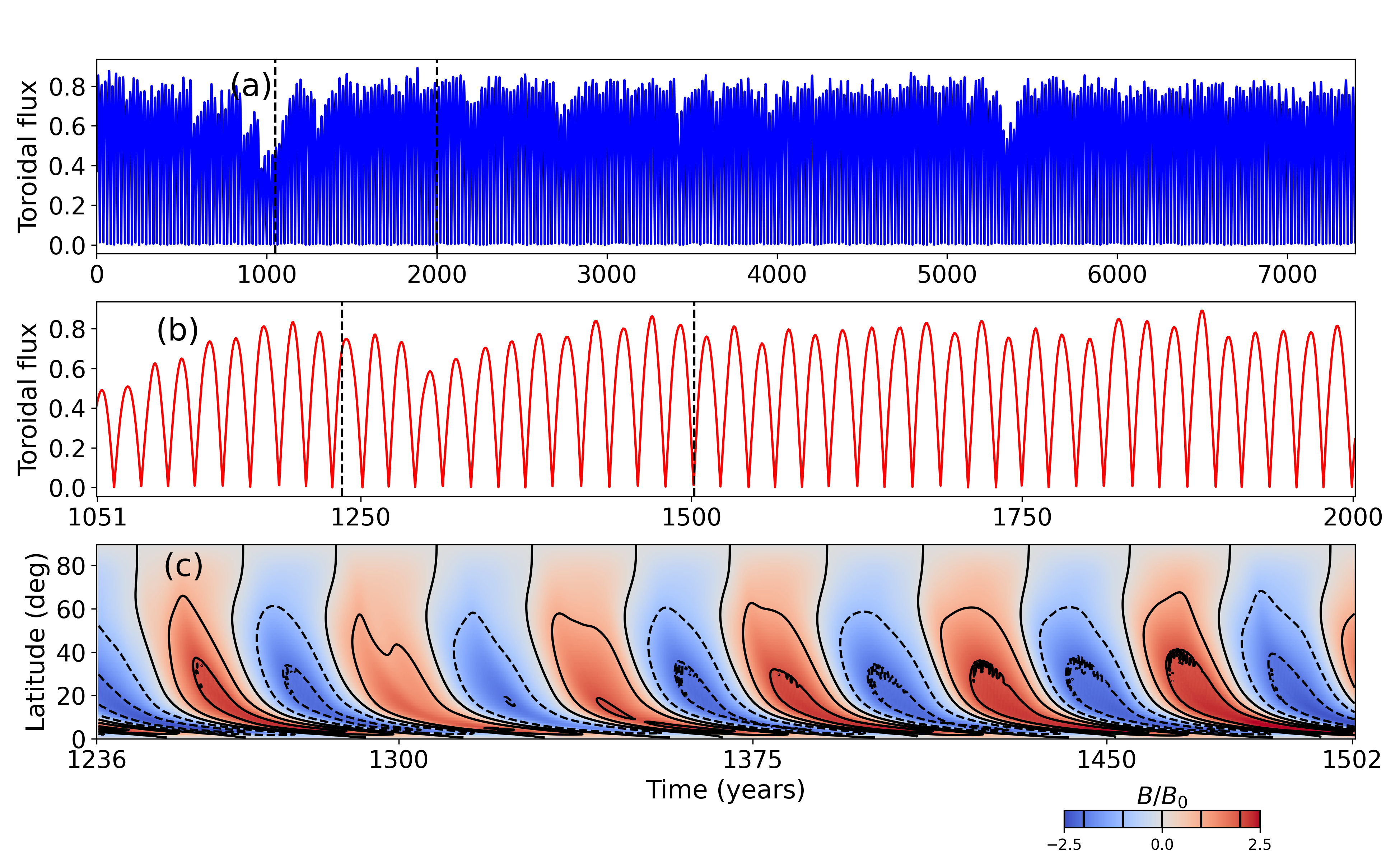}
    \caption{
    Same as \Fig{fig:app1_cr}, but from the super-critical regime.
    }
    \label{fig:app1_scr}
\end{figure}

In this section, we briefly highlight the salient features of the solar cycles obtained from Models I at near critical (\Fig{fig:app1_cr}) and super-critical regimes (\Fig{fig:app1_scr}). As seen from these two figures, the model shows irregular cycles with long-term modulation of the amplitude (top panels), solar-like oscillation of the magnetic field (middle panels), and the regular reversal of the polarity and the equatorward migration of the toroidal field at the base of the SCZ (bottom panels). We note that the cycle period is quite long ($\sim 22$ years) because of the weak meridional flow in this case. However, model II produces a reasonable cycle duration of about 12 years.  

\end{appendix}

\end{document}